\documentclass[aps,prd,twocolumn,groupedaddress,showpacs]{revtex4}

\usepackage{graphicx,dcolumn,bm,amssymb,amsmath,latexsym,amsfonts}

\begin{document}

%%%%%%%%%%%%%%%%%%%%%%%%%%%%%%%%%%%%%
%defining some commands
\newcommand{\nonu}{\nonumber}
\newcommand{\sm}{\small}
\newcommand{\noi}{\noindent}
\newcommand{\nl}{\newline}
\newcommand{\bc}{\begin{center}}
\newcommand{\ec}{\end{center}}
\newcommand{\be}{\begin{equation}}
\newcommand{\ee}{\end{equation}}
\newcommand{\beal}{\begin{align}}
\newcommand{\eeal}{\end{align}}
\newcommand{\bea}{\begin{eqnarray}}
\newcommand{\eea}{\end{eqnarray}}
\newcommand{\bnabla}{\mbox{\boldmath $\nabla$}}
\newcommand{\univec}{\textbf{a}}
\newcommand{\VectorA}{\textbf{A}}
\newcommand{\Pint}
%%%%%%%%%%%%%%%%%%%%%%%%%%%%%%%%%%%%

\title{Opposite charged two-body system of identical counter-rotating black holes}

\author{I. Cabrera-Munguia$^{1,}$\footnote{cabrera@zarm.uni-bremen.de}, Claus L\"ammerzahl$^{2,}$\footnote{laemmerzahl@zarm.uni-bremen.de}, L. A. L\'opez$^{3,}$\footnote{lalopez@uaeh.edu.mx}
and Alfredo Mac\'{\i}as$^{1,}$\footnote{amac@xanum.uam.mx}}
\affiliation{$^{1}$Departamento de F\'isica, Universidad Aut\'onoma Metropolitana-Iztapalapa A.P. 55-534, M\'exico D.F. 09340, M\'exico\\
$^{2}$ZARM, Universit\"{a}t Bremen, Am Fallturm, D-28359 Bremen, Germany\\
$^{3}$\'Area Acad\'emica de Matem\'aticas y F\'isica., UAEH, carretera Pachuca-Tulancingo km 4.5, C.P. 42184, Pachuca, Hidalgo, M\'exico}

%------------------------------Begin of the document --------------------------------

\begin{abstract}
A 4-parametric exact solution describing a two-body system of identical Kerr-Newman counter-rotating black holes endowed with opposite electric/magnetic charges is presented. The axis conditions are solved in order to really describe two black holes separated by a massless strut. Moreover, the explicit form of the horizon half length parameter $\sigma$ in terms of physical Komar parameters, i.e., Komar's mass $M$, electric charge $Q_{E}$, angular momentum $J$, and a coordinate distance $R$ is derived. Additionally, magnetic charges $Q_{B}$ arise from the rotation of electrically charged black holes. As a consequence, in order to account for the contribution to the mass of the magnetic charge, the usual Smarr mass formula should be generalized, as it is proposed by A. Tomimatsu, Prog. Theor. Phys. \textbf{72}, 73 (1984).

\end{abstract}

\maketitle

\section{INTRODUCTION}
Binary black hole systems in equilibrium, without a support strut in between, have been extensively studied in vacuum since the famous double-Kerr-NUT solution was presented by Kramer \emph{et al.} in 1980 \cite{KramerNeugebauer}. These type of solutions are nonregular \cite{Neugebauer}, due to the fact that at least one of the Komar masses results to be negative \cite{Hoenselaers,MRS}, appearing ring singularities off the axis. On the other hand, the electrovacuum sector has received less attention because the electromagnetic field increases considerably the difficulty of finding exact solutions in these type of two-body systems.

A binary system of identical Kerr-Newman (KN) sources separated by a massless strut (conical singularity) \cite{Israel} in between has been recently studied by Manko \emph{et al.} \cite{MRR}. The strut prevents the sources from falling onto each other and provides an interaction force which, nevertheless in this case, does not contain any spin-spin interaction. Furthermore, the equilibrium condition of the two-body system is reached after removing the strut and it reveals that the system is composed by identical counter-rotating relativistic disks, lying on the equatorial plane, whose individual electric charges, equal to their respective masses, result to have the same sign \cite{Parker,BM}. All of these aforementioned two-body systems do not contain individual magnetic charges; hence, they fulfill the standard Smarr formula for the mass \cite{Smarr}.

Additionally, following the ideas of Varzugin \cite{Varzugin}, we solved, in Ref. \cite{ICM} the axis conditions in order to define a 4-parametric asymptotically flat exact solution, which describes two unequal counter-rotating black holes separated by a massless strut. We established a straightforward procedure to obtain explicitly the functional form of the horizon of length $2\sigma$ in terms of Komar physical parameters \cite{Komar}. It determines the structure of the whole spacetime and its geometrical properties in a more physical way.

The main purpose of the present paper is to solve the axis conditions in order to describe a binary system of two identical counter-rotating black holes endowed with opposite electric/magnetic charges. We will show that magnetic charges arise as a result of the rotation of electrically charged black holes. In this description, in order to account for the contribution of the magnetic charge $Q_{B}$ to the mass, the Smarr mass formula \cite{Smarr} is generalized; it becomes a cubic equation. This modification is already proposed by Tomimatsu \cite{Tomimatsu1}, so that we only provide its physical form.

The interaction force related with the strut contains now, due to the rotation, a spin-spin interaction. Different limits of our solution are also discussed. Since the identical KN black holes are counter-rotating and have opposite electric charges, the full metric exhibits an equatorial antisymmetry property in the sense proposed by Ernst \emph{et al.} in \cite{EMR}. The upper black hole is characterized by having $J>0$ and $Q_{E}<0$, while the lower one has  $J<0$ and $Q_{E}>0$.

The outline of the paper is as follows: In Sec. II we solve the axis conditions for the case of a two-body system of identical counter-rotating black holes, endowed with opposite electromagnetic charges and separated by a conical line singularity \cite{Israel}. In Sec. III explicit formulas for the geometrical properties of the two-body system, and the analytical form of $\sigma$ in terms of Komar physical parameters, are derived. In Sec. IV the full metric for the extreme limit case is obtained. The concluding remarks are presented in Sec. V.

\section{OPPOSITE CHARGED TWO-BODY SYSTEM OF IDENTICAL COUNTER-ROTATING BLACK HOLES}
According to Ernst's formalism \cite{Ernst} the Einstein-Maxwell equations describing the stationary axisymmetric electrovacuum spacetimes can be reduced to the following system of equations:
\bea \begin{split}  \left({\rm{Re}} {\cal{E}}+|\Phi|^{2}\right)\Delta{\cal{E}}&=(\bnabla{\cal{E}}+
2\bar{\Phi}\bnabla \Phi)\bnabla {\cal{E}}, \\
 \left({\rm{Re}}{\cal{E}}+|\Phi|^{2}\right)\Delta \Phi&=(\bnabla{\cal{E}}+
2\bar{\Phi}\bnabla\Phi)\bnabla\Phi, \label{ERNST} \end{split} \eea

\noi where \bnabla\, and $\Delta$ are the gradient and Laplace operators defined in Weyl-Papapetrou cylindrical coordinates $(\rho,z)$ and acting over the complex potentials ${\cal{E}}=f - |\Phi|^{2} + i\Psi$ and $\Phi=-A_{4} +i A_{3}^{'}$. Here, $A_{4}$ is the electric potential and $A^{'}_{3}$ is associated with the magnetic potential $A_{3}$, both components of the electromagnetic 4-potential $A_{i}=(0,0,A_{3},A_{4})$. Any solution of
Eq.(\ref{ERNST}) determines the metric functions $\gamma$ and $\omega$ of the line element \cite{Papapetrou}
\be ds^{2}=f^{-1}\left[e^{2\gamma}(d\rho^{2}+dz^{2})+\rho^{2}d\varphi^{2}\right]- f(dt-\omega d\varphi)^{2},
\label{Papapetrou}\ee

\noi by means of the following set of differential equations:
\begin{widetext}
\bea \begin{split}
4\gamma_{\rho}&=\rho f^{-2} \left[|{\cal{E}}_{\rho}+
2\bar{\Phi}\Phi_{\rho}|^{2} -|{\cal{E}}_{z}+ 2\bar{\Phi}\Phi_{z}|^{2}\right]
- 4\rho f^{-1}(|\Phi_{\rho}|^{2}- |\Phi_{z}|^{2}),\\
2\gamma_{z}&=\rho f^{-2}{\rm{Re}} \left[({\cal{E}}_{\rho}+
2\bar{\Phi}\Phi_{\rho})(\bar{{\cal{E}}}_{z}+ 2\bar{\Phi}\Phi_{z})\right]
-4\rho f^{-1} {\rm{Re}(\bar{\Phi}_{\rho}\Phi_{z})},\\
\omega_{\rho}&=-\rho f^{-2}{\rm{Im}}( {\cal{E}}_{z}+ 2 \Phi\bar{\Phi}_{z}),\qquad
\omega_{z}=\rho f^{-2}{\rm{Im}}( {\cal{E}}_{\rho}+ 2 \Phi\bar{\Phi}_{\rho}),
\label{metricfunctions}\end{split}\eea

\noi where the subindices $\rho$ and $z$ denote partial differentiation, the bar over a symbol represents complex conjugation and $|x|^{2}=x \bar{x}$. An electrovacuum exact solution of Eq.(\ref{ERNST}), describing a binary system composed by KN sources, can be obtained with the aid of the Sibgatullin Method (SM) \cite{Sibgatullin,RMJ}. Following this approach, the Ernst potentials ${\cal{E}}, \Phi$ and the full metric read \cite{RMJ}
\bea \begin{split}
{\cal{E}}&=\frac{E_{+}}{E_{-}},\qquad \Phi=\frac{F}{E_{-}},\qquad f=\frac{D}{2 |E_{-}|^{2}}, \qquad \omega=\frac{ {2\rm {Im}}\left[E_{-}(\bar{G_{o}}+\bar{H_{o}})- F \bar{I} \right]}{D}
,\qquad e^{2\gamma}=\frac{D}{2|\mathfrak{a}_{+}|^{2} \prod_{n=1}^{4}r_n},\\
D&=E_{+}\bar{E}_{-}+\bar{E}_{+}{E}_{-}+2 F \bar{F},\qquad
E_{\pm}=\left|\begin{array}{ccccc}
1 & 1  & 1 & 1 & 1 \\
\pm1 & {} & {}& {} & {} \\
\pm1 & & {} C & {} & {}\\
0 & {} & {} & {} & {} \\
0 & {} & {} & {} & {} \\
\end{array}
\right|,\qquad
F=-\left|
\begin{array}{ccccc}
0 & f(\alpha_{1})  & f(\alpha_{2}) & f(\alpha_{3})& f(\alpha_{4}) \\
1 & {} & {}& {} & {} \\
1 & & {} C & {} & {}\\
0 & {} & {} & {} & {} \\
0 & {} & {} & {} & {} \\
\end{array}
\right|,\\
G_{o}&= \left|
\begin{array}{ccccc}
0 & p_{1} & p_{2} & p_{3} & p_{4} \\
1 & {} & {} & {} & {}  \\
1 & {} & C  \\
0 & {} & {} & {} & {} \\
0 & {} & {} & {} & {} \\
\end{array}
\right|,\qquad
H_{o}= \left|
\begin{array}{ccccc}
0 & 1 & 1 & 1 & 1 \\
-\beta_{1} & {} & {} & {} & {}  \\
-\beta_{2} & {} & C  \\
\bar{e_{1}} & {} & {} & {} & {} \\
\bar{e_{2}} & {} & {} & {} & {} \\
\end{array}
\right|,\qquad
I= \left|
\begin{array}{cccccc}
f_{1}+f_{2} & 0 & f(\alpha_{1}) & f(\alpha_{2}) & f(\alpha_{3}) & f(\alpha_{4})\\
z & 1 &  1 & 1 & 1 & 1  \\
-\beta_{1} & -1 & {} & {} & {}  \\
-\beta_{2} & -1 & {} & {}  & {} \\
\bar{e}_{1} & 0 & {} & C & {} & {} \\
\bar{e}_{2} & 0 & {} & {} & {} & {}\\
\end{array}
\right|,\\
C&=\left(\begin{array}{cccc}
\gamma_{11}r_{1} & \gamma_{12}r_{2} & \gamma_{13}r_{3}& \gamma_{14}r_{4} \\
\gamma_{21}r_1 & \gamma_{22}r_{2} & \gamma_{23}r_{3}& \gamma_{24}r_{4} \\
M_{11} & M_{12} & M_{13}& M_{14}\\
M_{21} & M_{22} & M_{23}& M_{24}\\
\end{array}
\right),\qquad
 \mathfrak{a}_{+}=\left|
\begin{array}{cccc}
\gamma_{11} & \gamma_{12} & \gamma_{13}& \gamma_{14} \\
\gamma_{21} & \gamma_{22} & \gamma_{23}& \gamma_{24} \\
M_{11} & M_{12} & M_{13}& M_{14}\\
M_{21} & M_{22} & M_{23}& M_{24}\\
\end{array}
\right|,\qquad p_{n}=2z-\alpha_{n}-r_{n},\\
M_{jn}&=\left[\bar{e}_{j}+2\bar{f}_{j} f(\alpha_{n})\right](\alpha_{n}-\bar{\beta}_{j})^{-1}, \qquad f(\alpha_{n})=\sum_{j=1}^{2} f_{j}\gamma_{jn},\qquad \gamma_{jn}=(\alpha_{n}-\beta_{j})^{-1},\qquad r_{n}=\sqrt{\rho^{2}+(z-\alpha_{n})^{2} }. \end{split} \label{generalsolution}\eea

\noi where the parameters $e_{j}$ are functions of $\alpha_{n}$, $f_{j}$, and $\beta_{j}$, they read:
\be e_{1}=\frac{2 \prod_{n=1}^{4}(\beta_{1}-\alpha_{n})}
{(\beta_{1}-\beta_{2})(\beta_{1}-\bar{\beta}_{1})(\beta_{1}-\bar{\beta}_{2})}-\sum_{k=1}^{2} \frac{2f_{1}\bar{f}_{k}}{\beta_{1}-\bar{\beta}_{k}},\qquad
e_{2}=\frac{2 \prod_{n=1}^{4}(\beta_{2}-\alpha_{n})}
{(\beta_{2}-\beta_{1})(\beta_{2}-\bar{\beta}_{1})(\beta_{2}-\bar{\beta}_{2})}-\sum_{k=1}^{2} \frac{2f_{2}\bar{f}_{k}}{\beta_{2}-\bar{\beta}_{k}}. \label{the-es} \ee
\end{widetext}

Equation (\ref{generalsolution}) contains a set of twelve algebraic parameters \{$\alpha_{n}$,$f_{j}$, $\beta_{j}$\}, where the real or complex values of $\alpha_{n}$ define subextreme objects (black holes) or hyperextreme objects (relativistic disks). It is important to note that the metric (\ref{generalsolution}) is not asymptotically flat at spatial infinity, since NUT sources \cite{NUT} as well as the total magnetic charge are present. Therefore, in order to get rid of such monopolar terms, which break the asymptotic flatness of the solution, it is necessary to impose and solve the corresponding conditions on the symmetry axis (axis conditions). By construction, Eq.(\ref{generalsolution}) satisfies an elementary flatness condition on the upper part of the symmetry axis: $\omega(\alpha_{1}<z<\infty)=0$ and $\gamma(\alpha_{1}<z<\infty)=0$. Besides, the metric function $\gamma$ ensures the fulfillment of the balance condition on the lower part of the symmetry axis: $\gamma(-\infty<z<\alpha_{4})=0$. The remaining conditions on the symmetry axis read
\bea  \begin{split} \omega(\rho=0,\alpha_{2}<z<\alpha_{3})&= 0, \\
\omega(\rho=0,-\infty<z<\alpha_{4})&= 0. \end{split} \label{twoconditions}\eea

A straightforward simplification leads us to the following algebraic system of equations:
\bea  \begin{split}  {\rm{Im}}[\mathfrak{\bar{a}}_{-}(\mathfrak{g}_{-}+\mathfrak{h}_{-})]&= 0, \qquad
{\rm{Im}}[\mathfrak{\bar{a}}_{+}(\mathfrak{g}_{+}+\mathfrak{h}_{+})]= 0,  \\
\mathfrak{g}_{\pm}&=\left|
\begin{array}{ccccc}
0 & 2 & 2 & 1\pm1 & 1\pm1 \\
1 & {} & {} & {} & {}  \\
1 & {} & (\mathfrak{a}_{\pm})  \\
0 & {} & {} & {} & {} \\
0 & {} & {} & {} & {} \\
\end{array}
\right|, \\
\mathfrak{h}_{\pm}&=\left|
\begin{array}{ccccc}
0 & 1 & 1 & 1 & 1 \\
1 & {} & {} & {} & {}  \\
1 & {} & (\mathfrak{a}_{\pm})  \\
\bar{e}_{1} & {} & {} & {} & {} \\
\bar{e}_{2} & {} & {} & {} & {} \\
\end{array}
\right|,\\
\mathfrak{a}_{\pm}&=\left|\begin{array}{cccc}
\pm \gamma_{11} & \pm \gamma_{12} & \gamma_{13}& \gamma_{14} \\
\pm \gamma_{21} & \pm \gamma_{22} & \gamma_{23}& \gamma_{24} \\
M_{11} & M_{12} & M_{13}& M_{14}\\
M_{21} & M_{22} & M_{23}& M_{24}\\
\end{array}
\right|.
\end{split} \label{algebraicequations}\eea

The total mass $\mathcal{M}$, total electric charge $\mathcal{Q}$, and total magnetic charge $\mathcal{B}$ of our binary system can be calculated asymptotically from the Ernst potentials on the symmetry axis, which lead to
\be {\rm{Re}}[e_{1}+e_{2}]=-2\mathcal{M}, \qquad f_{1}+f_{2}= \mathcal{Q}+ i\mathcal{B}.  \label{Total}\ee

Replacing Eq.(\ref{the-es}) into the first equation of (\ref{Total}) yields the relation
\be \beta_{1} + \beta_{2} + \bar{\beta}_{1} + \bar{\beta}_{2}=-2\mathcal{M}.\ee

\begin{figure}[ht]
\centering
\includegraphics[width=2cm,height=6cm]{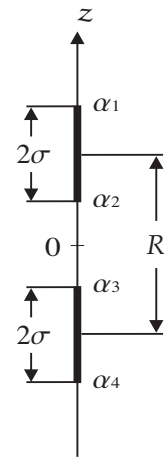}\\
\caption{Two identical KN black holes on the symmetry axis, with the values $\alpha_{1}=-\alpha_{4}=R/2+\sigma$,
$\alpha_{2}=-\alpha_{3}=R/2-\sigma$, and $R>2\sigma$. }
\label{DKNidentical}\end{figure}

By choosing $\beta_{1} + \beta_{2}=-\mathcal{M}:=-2M$, $\mathcal{Q}:=0$, and $\mathcal{B}:=0$, we are describing a system of two identical counter-rotating KN black holes of mass $M$, endowed with opposite electric/magnetic charge $Q_{E}/Q_{B}$ with a well-known conical line singularity \cite{Israel} in between. The corresponding horizons of the black holes are defined by the real values of the constant parameters $\alpha_{n}$, they fulfill the relation $\alpha_{1}+\alpha_{4}= \alpha_{2}+\alpha_{3}=0$, as shown in Fig. \ref{DKNidentical}. The parameters $\alpha_{n}$ can be written in terms of the relative coordinate distance $R$ and the half length $\sigma$ of each rod describing the black holes; they read
\be \alpha_{1}=-\alpha_{4}=\frac{R}{2}+\sigma, \qquad \alpha_{2}=-\alpha_{3}=\frac{R}{2}-\sigma.\ee

An explicit solution to the algebraic equations (\ref{algebraicequations}) is given by
\bea \begin{split} f_{1,2}&= \mp  \frac{q_{o}}{\sqrt{p+i\delta}},\quad
\beta_{1,2}= -M\pm\sqrt{p+i \delta},\\
p&=R^{2}/4-M^{2}+\sigma^{2}, \\
\delta&=\sqrt{(R^{2}-4M^{2})(M^{2}-\sigma^{2}-\mu Q_{o}^{2})},\\
q_{o}&:=Q_{o}(R/2-M),  \quad \mu:=\frac{R-2M}{R+2M}. \label{implicitsolution2}\end{split}\eea

\noi where the constant parameter $Q_{o}$ is the value of the electric charge $Q_{E}$ in absence of magnetic charge $Q_{B}$. It is important to note the fact that $M$, $J$, $-Q_{E}$ are the parameters characterizing the upper constituent of the system, while $M$, $-J$, $+Q_{E}$ characterize the lower part of the system. The black holes are separated by a coordinate distance $R$. In order to guarantee equatorial antisymmetry of the exact solution \cite{EMR}, the electric and magnetic charges of the constituents should have opposite sign. By using Eq.(\ref{implicitsolution2}), one is able to prove that Eq.(\ref{generalsolution}) reduce to

\begin{widetext}
\bea \begin{split}
{\cal{E}}&=\frac{\Lambda-\Gamma}{\Lambda+\Gamma},\quad \Phi=\frac{\chi}{\Lambda+\Gamma},\quad
f=\frac{|\Lambda|^{2}-|\Gamma|^{2}+|\chi|^{2}}{| \Lambda+\Gamma|^{2}}, \quad \omega=\frac{{\rm{Im}}\left[(\Lambda+\Gamma)\bar{\mathcal{G}}-\chi\bar{\mathcal{I}}\right]}{|\Lambda|^{2}-
|\Gamma|^{2}+|\chi|^{2}},\quad e^{2\gamma}=\frac{|\Lambda|^{2}-|\Gamma|^{2}+|\chi|^{2}}{\kappa_{o}^{2} r_{1}r_{2}r_{3}r_{4}},\\
\Lambda&=4\sigma^{2}(\kappa_{+}+2q_{o}^{2})(r_{1}-r_{3})(r_{2}-r_{4})+ R^{2}(\kappa_{-}-2q_{o}^{2})(r_{1}-r_{2})(r_{3}-r_{4})\\
&+ 2\sigma R(R^{2}-4\sigma^{2})
\left[\sigma R(r_{1}r_{4}+r_{2}r_{3})+i\delta(r_{1}r_{4}-r_{2}r_{3})\right],\\
\Gamma&=2M\sigma R(R^{2}-4\sigma^{2})[ \sigma R(r_{1}+r_{2}+r_{3}+r_{4})
-(2M^{2}-i\delta)(r_{1}-r_{2}-r_{3}+r_{4})],\\
\chi&=4q_{o}\sigma R[(R-2\sigma)(\epsilon_{+}+4M^{2})(r_{1}-r_{4})+(R+2\sigma)(\epsilon_{-}-4M^{2})(r_{2}-r_{3})], \\
\mathcal{G}&=-2 z \Gamma + 2\sigma R[ 4\sigma \kappa_{+}(r_{1}r_{2}-r_{3}r_{4})  +2R\kappa_{-}(r_{1}r_{3}-r_{2}r_{4})-M(R-2\sigma)\nu_{+}(r_{1}-r_{4})
-M(R+2\sigma)\nu_{-}(r_{2}-r_{3})  ], \\
\mathcal{I}&= 4M q_{o}[2\sigma^{2}(R^{2}-4M^{2}-2i\delta)(r_{1}r_{2}+r_{3}r_{4})
+ R^{2}(2M^{2}-2\sigma^{2}+i\delta) (r_{1}r_{3}+r_{2}r_{4})]-2q_{o}(R^{2}-4\sigma^{2})\\
&\times\left\{ 2M\left[(\epsilon_{+}+4M^{2})r_{1}r_{4}-(\epsilon_{-}-4M^{2})r_{2}r_{3}\right]
+ \sigma R\left[(\epsilon_{+}+8M^{2})(r_{1}+r_{4})+(\epsilon_{-}-8M^{2})(r_{2}+r_{3})+8\sigma M R\right]\right\}, \\
\kappa_{o}&:=4\sigma^{2}R^{2}(R^{2}-4\sigma^{2}),\quad \kappa_{\pm}:=M^{2}(R^{2}-4\sigma^{2})\pm 2q_{o}^{2}, \quad
\nu_{\pm}:=\epsilon_{\pm}(R\pm2\sigma)^{2}\pm 8q_{o}^{2},\quad \epsilon_{\pm}:=\sigma R\mp(2M^{2}-i \delta), \label{4-parametros}  \end{split} \eea
\end{widetext}

\noi where $r_{n}$ have the following reparametrized form:
\bea \begin{split} r_{1,2}&=\sqrt{\rho^{2}+\left(z-R/2 \mp \sigma\right)^{2}}, \\
r_{3,4}&=\sqrt{\rho^{2}+\left(z+R/2 \mp \sigma\right)^{2}}. \end{split}\eea

The corresponding Ernst potentials on the symmetry axis are given by
\bea \begin{split}
e(z)&=\frac{e_{+}}{e_{-}},\qquad f(z)=\mp\frac{2|q_{o}| }{e_{-}}, \\
 e_{\pm}&=z^{2} \mp 2M z + 2M^{2}-R^{2}/4-\sigma^{2}-i\delta.\end{split}\eea

The above solution [Eq.(\ref{4-parametros})] possesses the equatorial antisymmetry property in the sense of \cite{EMR}, according to which after making the change $z\rightarrow -z$, the metric function $\omega$ changes its global sign and the Ernst potentials on the symmetry axis satisfy the relations $e(z)e(-z)=1$ and $f(z)=\mp f(-z)e(z)$. From Eq.(\ref{4-parametros}), $Q_{o}=0$ defines a two-body system of equal counter-rotating black holes \cite{Varzugin}. It is a particular case of the solution presented in Ref. \cite{ICM}. Furthermore, if $\sigma=\sqrt{M^{2}-\mu Q_{o}^{2}}$ with $Q_{o}=-Q_{E}$, Eq.(\ref{4-parametros}) represents a binary system of identical opposite charged Reissner-Nordstr\"{o}m black holes \cite{Varzugin1}. On the other hand, the analysis of the energy-momentum tensor of the strut in between, leads us to the expression for the interaction force \cite{Israel,Weinstein}:
\be \mathcal{F}=\frac{1}{4}(e^{-\gamma_{0}}-1)= \frac{M^{2}(R^{2}-4\sigma^{2})+Q_{o}^{2}(R-2M)^{2}}{(R^{2}-4M^{2})(R^{2}-4\sigma^{2}-4\mu Q_{o}^{2})},\label{force}\ee
\noi where $\gamma_{0}$ is the value of the metric function $\gamma$ on the region of the strut.

\section{ANALYTIC FORM OF $\sigma$ AND GEOMETRICAL PROPERTIES OF THE SOLUTION }
In this case, the half length of the horizon $\sigma$ can be written as a function of physical Komar parameters \cite{Komar}: $M$, $Q_{E}$, $J$, and a coordinate distance $R$. Nevertheless, the individual magnetic charge $Q_{B}$ is not vanishing in this approach. In order to calculate $\sigma$ we are going to use the Tomimatsu's formulas \cite{Tomimatsu0},
\bea \begin{split} M&=-\frac{1}{8\pi}\int_{H} \omega \Psi_{z}\, d\varphi dz, \\
Q_{E}&=\frac{1}{4\pi}\int_{H}\omega A_{3z}^{'}\, d\varphi dz, \quad Q_{B}=\frac{1}{4\pi}\int_{H}\omega A_{4z} \, d\varphi dz, \\
J&=-\frac{1}{8\pi}\int_{H}\omega\left[1+\frac{1}{2}\omega \Psi_{z}
-\tilde{A}_{3}A_{3z}^{'}-(A_{3}^{'}A_{3})_{z}\right]d\varphi dz , \end{split}\label{Tomi}\eea

\noi with $\tilde{A}_{3}:=A_{3}+ \omega A_{4}$ and the magnetic potential $A_{3}$ is defined as the real part of Kinnersley's potential $\Phi_{2}$ \cite{Kinnersley}. Using the SM \cite{RMJ}, one finds that
\be \Phi_{2}=-i\frac{I}{E_{-}}=i \left( z \Phi- \frac{\mathcal{I}}{\Lambda+\Gamma}\right).\ee

Since the black holes are identical, the horizon for the upper object is defined as a null hypersurface
$H=\{-\sigma\leq z - \frac{R}{2}\leq \sigma,\,0 \leq \varphi \leq 2\pi,\, \rho\rightarrow 0\}$. From Eq.(\ref{Tomi}), one can show that $M$ represents the individual mass of each of the black holes. On the other hand, the electric and magnetic charges read
\bea \begin{split}
Q_{E}&=-\frac{Q_{o}(R^{2}-4M^{2})}{R^{2}-4\sigma^{2}-4\mu Q_{o}^{2}},\\
Q_{B}&=\frac{2Q_{o}\sqrt{(R^{2}-4M^{2})(M^{2}-\sigma^{2}-\mu Q_{o}^{2})}}{
R^{2}-4\sigma^{2}-4\mu Q_{o}^{2}}.\label{charges}\end{split}\eea

\noi Notice that the electric and magnetic charges possess opposite sign. After combining both equations in (\ref{charges}), one gets
\be Q_{E}^{2}+Q_{B}^{2}=-Q_{E}Q_{o}, \qquad (Q_{E}<0, \, Q_{B}>0),\ee

\noi where the parameter $Q_{o}$ allows us to define a new auxiliary variable as follows:
\be Q_{o}=-Q_{E}X, \qquad X:= 1+\frac{Q_{B}^{2}}{Q_{E}^{2}}, \label{theX}\ee

\noi and $\sigma$ can be written as a function of $X$ in the form
\be \sigma=\sqrt{X(M^{2}-Q_{E}^{2} \mu X)+\frac{R^{2}}{4}\left(1-X\right)}. \label{sigma} \ee

The last integral in Eq.(\ref{Tomi}), which defines the angular momentum $J$, is not vanishing and the usual Smarr formula for the mass \cite{Smarr} is not anymore fulfilled. As Tomimatsu proposed \cite{Tomimatsu1}, it should be enhanced in order to include the contribution to the mass of the magnetic charge; it acquires the corresponding additional term:
\be M=\frac{\kappa S}{4\pi} +2 \Omega J +\Phi^{H}_{E} Q_{E}+ M_{A}^{S} =\sigma +2 \Omega J +\Phi^{H}_{E} Q_{E}+ M_{A}^{S} \label{newSmarr}\ee

\noi where $\Phi^{H}_{E}=-A_{4}^{H}-\Omega A_{3}^{H}$ is the electric potential in the frame rotating
with the black hole and $M_{A}^{S}$ is an extra boundary term associated to the magnetic charge \cite{Tomimatsu1}, given by
\bea \begin{split}
M_{A}^{S}&=-\frac{1}{4\pi}\int_{H} \left(A_{3}A_{3}^{'}\right)_{z}\, d\varphi dz \\
&=2Q_{o}^{2}(R-2M)^{2}\left[\frac{M^{2}-\sigma^{2}-\mu Q_{o}^{2}}
{(R^{2}-4\sigma^{2}-4 \mu Q_{o}^{2})^{2}}\right]\\
&\times \left[\frac{(R+2\sigma)(R+4M+2\sigma)+4 \mu Q_{o}^{2}}
{2M(M+\sigma)(R+2\sigma)-\mu Q_{o}^{2}(R-2M)}\right]. \label{MAS} \end{split}\eea

Let us call $\omega^{H}$ the constant value of the metric function $\omega$ at the horizon. $\Omega:=1/\omega^{H}$ is the angular velocity. A simple calculation leads us to the following expressions for $\Omega$ and $\Phi^{H}_{E}$:
\bea \begin{split}
\Omega&=\frac{(R+2\sigma)\sqrt{\mu(M^{2}-\sigma^{2}-\mu Q_{o}^{2})}}{2M(M+\sigma)
(R+2\sigma)-\mu Q_{o}^{2}(R-2M)},\\
\Phi^{H}_{E}&= -\frac{Q_{o}(R-2M)(M+\sigma)}{2M(M+\sigma)
(R+2\sigma)-\mu Q_{o}^{2}(R-2M)}.\label{vel}\end{split}\eea

Replacing $Q_{o}$ from Eq.(\ref{theX}) and $\sigma$ from Eq.(\ref{sigma}) into Eqs.(\ref{MAS}) and (\ref{vel}), one gets
\bea \begin{split}
M_{A}^{S}&=\frac{\mu Q_{E}^{2}(X-1)[2(R+2\sigma)-(R-2M)X]}{2\left(M[R+2\sigma-(R-2M)X]-\mu Q_{E}^{2}X^{2}\right)}, \\
\Omega&=\frac{\mu}{2}\frac{(R+2\sigma)\sqrt{X-1}}{M[R+2\sigma-(R-2M)X]-\mu Q_{E}^{2}X^{2}}, \\
\Phi^{H}_{E}&= \frac{Q_{E}\mu(M+\sigma)X}{M[R+2\sigma-(R-2M)X]-\mu Q_{E}^{2}X^{2}}.
\label{newgeometrical} \end{split}\eea

Combining Eq.({\ref{newgeometrical}}) with each other, it is easy to find a kind of enhanced form for the Smarr formula, which includes the contribution to the mass of the boundary term associated with the magnetic charge \cite{Tomimatsu1},
\be M=\sigma +\Omega \left[2J-Q_{E}Q_{B}\left(1-\frac{Q_{B}^{2}}{Q_{E}^{2}}\right) \right] +\Phi^{H}_{E}\left(1+\frac{Q_{B}^{2}}{Q_{E}^{2}}\right)Q_{E}. \label{newSmarr}\ee

The substitution of $\sigma$ from Eq.(\ref{sigma}) into Eq.(\ref{newSmarr}) leads us to the following cubic equation in terms of the auxiliary variable $X$:
\be (X-1)\left[X-2\left(1-\frac{2M^{2}}{Q_{E}^{2}(1-\mu)}\right)\right]^{2}-\frac{4J^{2}}{Q_{E}^{4}}=0. \label{cubic}\ee

The explicit real root solution of Eq.(\ref{cubic}) is given by
\bea \begin{split}
X&=1+ \frac{[a+ [b-a^{3} +\sqrt{b(b-2a^{3})}]^{1/3}]^{2}}{[b-a^{3} +\sqrt{b(b-2a^{3})}]^{1/3}},\\
a&:=\frac{1}{3}\left(1-\frac{4M^{2}}{Q_{E}^{2}(1-\mu)}\right),\quad b:=\frac{2J^{2}}{Q_{E}^{4}},\quad b\geq2a^{3}. \label{solutionX} \end{split}\eea

From the second Eq.(\ref{theX}) the individual magnetic charge reads
\be Q_{B}=-Q_{E}\sqrt{X-1}. \label{magneticcharge}\ee

\noi In the electrostatic limit, i.e., $J=0$, $X=1$, $\sigma$ reduces to
\be \sigma_{E}=\sqrt{M^{2}-Q_{E}^{2} \mu }.\ee

\noi which is one special case of the corresponding relation given in \cite{Varzugin1}. On the other hand, in the vacuum limit, i.e., $Q_{E}=0$, $X=1+(1-\mu)^{2}J^{2}/4M^{4}$, $\sigma$ reads \cite{Varzugin}
\be \sigma_{V}=\sqrt{M^{2}-\frac{J^{2}}{M^{2}} \mu }.\ee

In both limits the magnetic charge, Eq.(\ref{magneticcharge}), vanishes. However, this is not true in the electrovacuum case as we shall show later. The interaction force due to the strut in between reads
\be \mathcal{F}=\frac{M^{2}}{R^{2}-4M^{2}}+(Q_{E}^{2}+Q_{B}^{2}) \mu  \left(\frac{R}{R^{2}-4M^{2}}\right)^{2}.\label{forceX} \ee

\noi Notice the explicit appearance of the magnetic charge.

It is worthwhile to mention that Eq.(\ref{forceX}) has the same form in the non-extreme case, as well as in the extreme case. The difference is that in the extreme case the condition $\sigma=0$ relates the parameters appearing in Eqs.(\ref{sigma}) and (\ref{cubic}).

\vspace{-0.3cm}\subsection{Analytic calculation of $\sigma$}\vspace{-0.2cm}
A counter-rotating opposite charged two-body system clearly reveals the appearance of magnetic charges as a consequence of the rotation of electrically charged black holes. In order to derive an explicit form of $\sigma$ we will consider first a system of two identical counter-rotating black holes in a weak electromagnetic field. Therefore, the corresponding value of $X$ is
\be X \simeq 1+\frac{J^{2}(1-\mu)^{2}}{4M^{4}}.\ee

Hence, the magnetic charge reads
\be Q_{B}\simeq -Q_{E} \frac{J(1-\mu)}{2M^{2}},\label{magnetic1}\ee

\noi and $\sigma$ reduces to
\be \sigma \simeq \sqrt{M^{2}-\frac{J^{2}}{M^{2}} \mu-Q_{E}^{2}\left(1+\frac{J^{2}(1-\mu)^{2}}{4M^{4}}\right)^{2}\mu}.\label{sigma1}\ee

\noi The interaction force can now be written as
\bea \begin{split}
\mathcal{F}& \simeq \frac{M^{2}}{R^{2}-4M^{2}}\\
&+Q_{E}^{2} \mu  \left(\frac{R}{R^{2}-4M^{2}}\right)^{2}\left(1+\frac{J^{2}(1-\mu)^{2}}{4M^{4}}\right).
\label{forcecase1} \end{split}\eea

Let us now consider a system of opposite electric charged black holes with slow rotation. The corresponding value of $X$ is now given by
\be X\simeq 1+\frac{4J^{2}(1-\mu)^{2}}{[4M^{2}-Q_{E}^{2}(1-\mu)]^{2}},\ee

\noi the magnetic charge reads
\be Q_{B} \simeq -Q_{E} \frac{2J(1-\mu)}{4M^{2}-Q_{E}^{2}(1-\mu)}, \label{magnetic2}\ee

\noi and $\sigma$ can be written as
\be \sigma \simeq \sqrt{M^{2}-Q_{E}^{2}\mu-8J^{2}\left(\frac{2M^{2}+Q_{E}^{2}(1-\mu)^{2}}
{[4M^{2}-Q_{E}^{2}(1-\mu)]^{2}}\right)\mu}.\label{sigma2}\ee

\noi The corresponding interaction force is given by
\bea \begin{split}
\mathcal{F}\simeq &\frac{M^{2}}{R^{2}-4M^{2}}+ Q_{E}^{2} \mu  \left(\frac{R}{R^{2}-4M^{2}}\right)^{2}\\
&\times \left(1+\frac{4J^{2}(1-\mu)^{2}}{[4M^{2}-Q_{E}^{2}(1-\mu)]^{2}}\right).
\label{forcecase2} \end{split}\eea

It is important to note that the interaction force corresponding to the two examples given above presents a
spin-spin interaction and consequently magnetic charges appear in the systems under consideration.

\vspace{-0.3cm} \subsection{Geometrical properties}\vspace{-0.2cm}
The surface gravity $\kappa$ and area of the horizon $S$ can be obtained directly from Eq.(\ref{4-parametros}) and without any previous knowledge of the explicit form of $\sigma$. In order to calculate $\kappa$, one can use the formula \cite{Tomimatsu1}
\be \kappa=\sqrt{-\Omega^{2}e^{-2\gamma^{H}}},\ee

\noi where $\gamma^{H}$ is the metric function $\gamma$ evaluated at the horizon. A straightforward calculation leads us to the following expressions for the surface gravity and the area of the horizon:
\bea \begin{split}
\kappa&=\frac{R \sigma(R+2\sigma)}{2M(M+\sigma)(R+2\sigma)(R+2M)-Q_{o}^{2}(R-2M)^{2}}\\
S&=4\pi \left[ 2M(M+\sigma)\left(1+\frac{2M}{R}\right)- \frac{Q_{o}^{2}(R-2M)^{2}}{R(R+2\sigma)}\right], \end{split}\eea

\noi where $Q_{o}$ was already defined in Eq.(\ref{theX}). One should note that the strut between the KN black holes
disappears in the limit $R\rightarrow \infty$ and the bodies are isolated. In this limit both magnetic charges vanish as well as the extra boundary term given by Eq.(\ref{MAS}) and therefore Eqs.(\ref{sigma1}) and (\ref{sigma2}) reduce to $\sigma=\sqrt{M^{2}-Q_{E}^{2}-J^{2}/M^{2}}$. In addition, if $R\rightarrow 2M$, the two horizons can touch each other, the angular velocities are stopped and the two-body system evolves into one single Schwarzschild black hole.

\vspace{-0.50cm}\section{The extreme limit of the solution}
The extreme limit can be obtained by setting $\sigma=0$ in Eq.(\ref{4-parametros}), and after a careful use of the l'H\^{o}pital's rule, one gets
\begin{widetext}
\bea \begin{split} {\cal{E}}&=\frac{\Lambda-2\alpha M x\Gamma_{+}}{\Lambda+2\alpha M x\Gamma_{+}},\quad \Phi=\frac{2q_{o}y\Gamma_{-}}{\Lambda+2\alpha M x\Gamma_{+}},\quad f=\frac{D}{N}, \quad
\omega=\frac{4\alpha^{2}\delta_{o}\,y(x^{2}-1)(y^{2}-1)W}{D}, \quad
e^{2\gamma}=\frac{D}{\alpha^{8}(x^{2}-y^{2})^{4}}, \\
\Lambda&= \alpha^{2}(\alpha^{2}-M^{2})(x^{2}-y^{2})^{2}+ \alpha^{2}M^{2}(x^{4}-1)
+ q_{o}^{2}(1-y^{4}) +2i\alpha^{2} \delta_{o}(x^{2}+y^{2}-2 x^{2} y^{2}), \\
\Gamma_{\pm}&= \pm \left\{\left(\sqrt{M^{2}-\mu Q_{o}^{2}}\mp i\sqrt{\alpha^{2}-M^{2}}\right)\left[\sqrt{M^{2}-\mu Q_{o}^{2}}(x^{2}-1) \pm i\sqrt{\alpha^{2}-M^{2}}(x^{2}-y^{2})\right]+ \mu Q_{o}^{2}(x^{2}-1)\right\}, \\
D&= [\alpha^{2}(\alpha^{2}-M^{2})(x^{2}-y^{2})^{2}+\alpha^{2}M^{2}(x^{2}-1)^{2} -q_{o}^{2}(y^{2}-1)^{2}]^{2}-16\alpha^{4}\delta_{o}^{2}x^{2} y^{2}(x^{2}-1)(1-y^{2}), \\
N&=\{ \alpha^{2}(\alpha^{2}-M^{2})(x^{2}-y^{2})^{2}+ \alpha^{2}M^{2}(x^{4}-1)+q_{o}^{2}(1-y^{4})
+ 2\alpha Mx[(\alpha^{2}-M^{2})(x^{2}-y^{2})+M^{2}(x^{2}-1)]\}^{2}\\
&+4\alpha^{2}\delta_{o}^{2}\left[\alpha(x^{2}+y^{2}-2x^{2}y^{2})+M x(1-y^{2})\right]^{2}, \\
W&= M\alpha^{2} [(\alpha^{2}-M^{2})(x^{2}-y^{2})(3x^{2}+y^{2})+ M^{2}(3x^{4}+6x^{2}-1) +8\alpha M x^{3}]
+ q_{o}^{2}[M(y^{2}-1)^{2}-4\alpha x y^{2}], \\
\delta_{o}&:=\sqrt{(\alpha^{2}-M^{2})(M^{2}-\mu Q_{o}^{2})},\qquad \alpha:=\frac{R}{2}, \label{extreme}\end{split}\eea

\noi where $(x,y)$ are prolate spheroidal coordinates
\be x=\frac{r_{+}+r_{-}}{2\alpha}, \qquad y=\frac{r_{+}-r_{-}}{2\alpha}, \qquad r_{\pm}=\sqrt{\rho^{2} +(z\pm\alpha)^{2}}, \label{cilindricas}\ee

\noi related to the cylindrical coordinates $(\rho,z)$ through the following transformation formulas:
\be \rho=\alpha\sqrt{(x^{2}-1)(1-y^{2})}, \qquad z=\alpha xy. \label{cilindricas} \ee

The extreme limit case given by Eq.(\ref{extreme}) is a 3-parametric exact solution where the physical parameters
are related by Eq.(\ref{cubic}) and by the following equation:
\be X^{2}+ \frac{(\alpha+M)^{2}}{Q_{E}^{2}}X-\frac{\alpha^{2}}{Q_{E}^{2}}\left(\frac{\alpha+M}{\alpha-M}\right)=0. \label{Xextreme}\ee

Nevertheless, it is quite complicated to derive an analytic expression of one of the parameters in terms of the other three. After combining Eqs.(\ref{cubic}) and (\ref{Xextreme}), it is possible to get the following relation:
\bea \begin{split}
|J|&=\frac{\left[2\mu (M^{2}(1-2\mu)-Q_{E}^{2}(1-\mu)^{2})+M \sqrt{\mu \left[4M^{2}\mu+Q_{E}^{2}(1-\mu^{2})^{2}\right]}\right]}{2|Q_{E}|\mu^{3/2}(1-\mu)^{3}}\\
&\times \sqrt{M \sqrt{\mu \left[4M^{2}\mu+Q_{E}^{2}(1-\mu^{2})^{2}\right]}-\mu(2M^{2}+Q_{E}^{2}(1-\mu)^{2})},
\label{angularmomentumextreme} \end{split}\eea
\end{widetext}

\noi whose asymptotic expansion lead us to
\be \frac{|J|}{M \sqrt{M^{2}-Q_{E}^{2}}}\simeq 1 +\frac{2M^{4}-Q_{E}^{2}(M^{2}-Q_{E}^{2})}{M (M^{2}-Q_{E}^{2})}\left(\frac{1}{R}\right)>1, \ee

\noi and it implies that the inequality $J^{2}/M^{2}>M^{2}-Q_{E}^{2}>0$ holds for positive values of the distance
$R \gg 2M$, for which $0<\mu<1$. The equality $J^{2}/M^{2}=M^{2}-Q_{E}^{2}$ is reached if the distance becomes large enough and tends to infinity (i.e, $\mu=1$), where the black holes are isolated. It is important to stress the fact that magnetic charges depend also on the coordinate distance $R$ as shown in Eqs.(\ref{magnetic1}) and (\ref{magnetic2}), but they vanish if the distance tends to infinity.\nl

It should be pointed out that the metric Eq.(\ref{extreme}) fulfills the axis condition for all the regions on the symmetry axis: $\omega(y=\pm1)=0$ for $|z|>\alpha$ and $\omega(x=1)=0$ for $|z|<\alpha$. It reduces to well-known limits: by setting $Q_{E}=0$ it results to be one particular case of the Kinnersley-Chitre solution \cite{KCH}.
The extreme double-Reissner-Nordstr\"{o}m (DRN) solution is obtained if $|Q_{E}|=\mu^{-1/2}M > M$ and $J=0$ (black dihole solution \cite{Emparan}). The extreme DRN solution was already considered in Ref. \cite{Cabrera} for unequal constituents. However, the expression for the interaction force between identical extreme Reissner--Nordstr\"{o}m components is not written, it explicitly reads
\be F=\frac{2M^{2}}{R^{2}-4M^{2}} \left[1+ \frac{2M^{2}}{R^{2}-4M^{2}}\right], \quad R>2M. \label{extremeDRN}\ee

\section{CONCLUDING REMARKS}
This paper deals with a complementary asymptotically flat exact solution related to identical counter-rotating black hole sources, in the presence of electromagnetic field. Particularly, the black holes are endowed with opposite electric/magnetic charges. Our description provides an analytical way to derive the expression for $\sigma$ in terms of the physical Komar parameters and the coordinate distance. This new exact solution gives a physical explanation of the appearance of magnetic charges in the solution as a consequence of the rotation of electrically charged black holes. Moreover, the presence of magnetic charges violates the usual Smarr formula for the mass; it should be enhanced in order to take into account the contribution to the mass of the magnetic charge.\nl

\noi On the other hand, it is worthwhile to mention that according to the positive mass theorem \cite{SchoenYau1,SchoenYau2} a regular solution of Eqs.(\ref{4-parametros}) and (\ref{extreme}) should fulfill the mass formula Eq.(\ref{newSmarr}): $M\geq \Omega[2J-Q_{E}Q_{B}(1-Q_{B}^{2}/Q_{E}^{2})] +\Phi^{H}_{E}(1+Q_{B}^{2}/Q_{E}^{2})Q_{E}>0$. Nevertheless, the theorem does not imply that the condition $M>0$ is enough to prove regularity of the solutions. Hence, one has to look at the denominator of the Ernst potentials in order to prevent such singularities. Nowadays, a reliable analytical study of singularities does not exist and it probably is due to the high order polynomials appearing in the denominator of the Ernst potentials. This inconvenience leads us to resort to numerical analysis. Let us look for such singularities in the extreme limit case, we have that
\be \Lambda+2\alpha M x \Gamma_{+}=F_{R}+ i F_{I}=0, \label{ernstsingular}\ee

\noi where
\bea \begin{split} F_{R}&= \alpha^{2}(\alpha^{2}-M^{2})(x^{2}-y^{2})^{2}+ \alpha^{2}M^{2}(x^{4}-1)\\
& + 2\alpha Mx\left[(\alpha^{2}-M^{2})(x^{2}-y^{2})+M^{2}(x^{2}-1)\right]  \\
& + q_{o}^{2}(1-y^{4})=0,\\
F_{I}&=2\alpha\delta_{o}\left[\alpha(x^{2}+y^{2}-2x^{2}y^{2})+M x(1-y^{2})\right]=0. \end{split}\label{curves} \eea

In Weyl-Papapetrou cylindrical coordinates $(\rho,z)$ the interior naked singularity of a black hole lies on the
symmetry axis. In the plane $(x,y)$ the region $x\geq1$, $|y|\geq1$, contains the allowed values for the solution Eq.(\ref{extreme}) in order to avoid naked singularities off the axis. The curves defined by Eq.(\ref{curves}) do not have intersections in such region if $M>0$ and therefore the solution is regular (see Fig. 2). Moreover, if $M<0$, in the solution Eq.(\ref{extreme}) ring singularities off the axis arise due to the intersections of the curves of Eq.(\ref{curves}) in the region $x>1$, $|y|<1$, (see Fig. 3).
\begin{figure}[ht]
\begin{minipage}{0.49\linewidth}
\centering
\includegraphics[width=4.25cm,height=5cm]{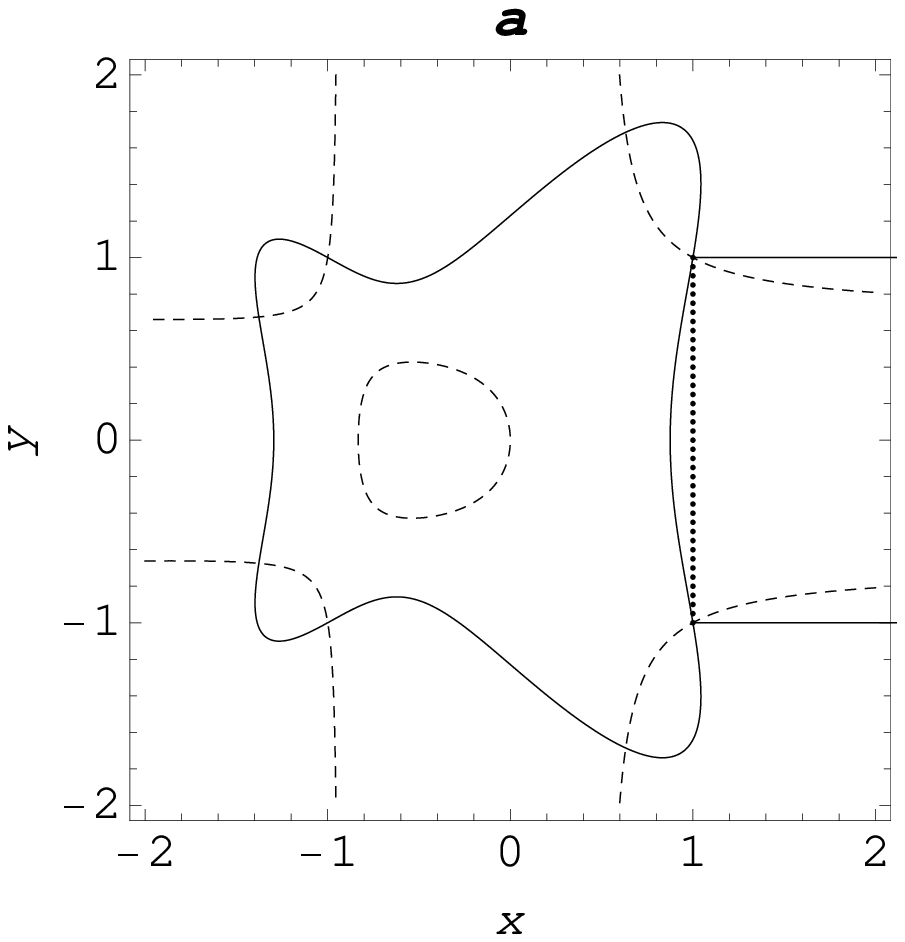}
%\center{a)} opcional
\end{minipage}
\begin{minipage}{0.49\linewidth}
\centering
\includegraphics[width=4.25cm,height=5cm]{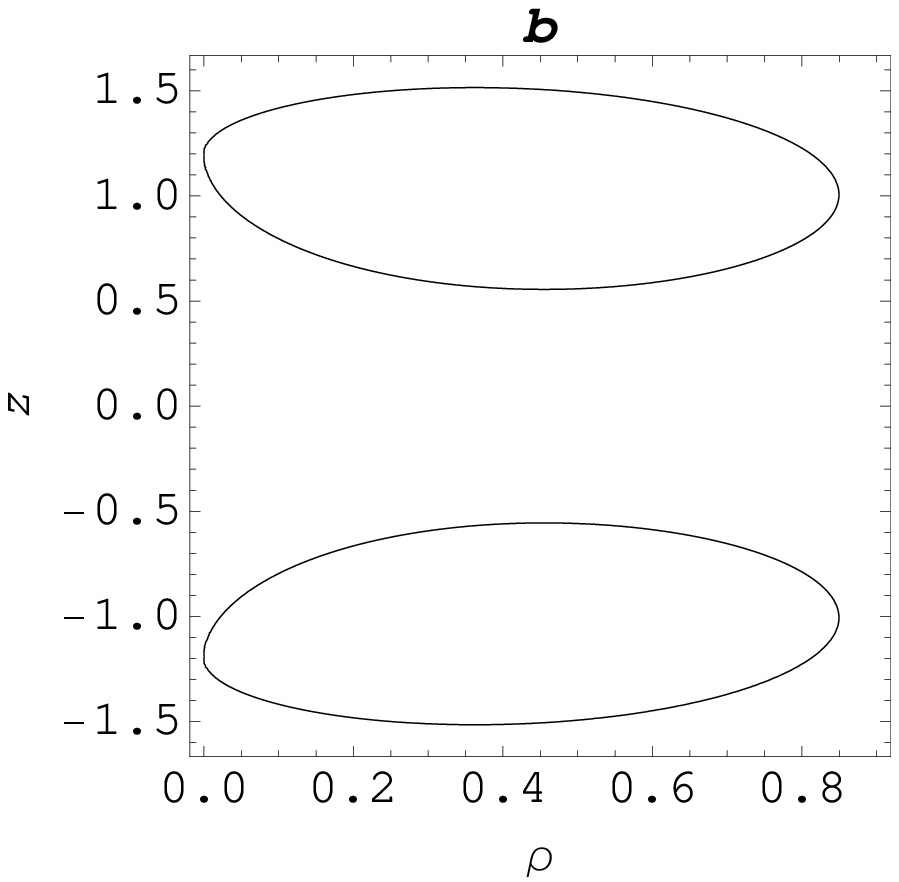}
%\center{b)} opcional
\end{minipage}
\caption{(a) No zeros in the denominator of the Ernst potentials in the $(x,y)$ plane, for the values $M=1$, $Q_{E}=-0.1$, $Q_{B}=0.15$, $J=3.31$, and $\alpha=1.2$. $F_{R}$ and $F_{I}$ are represented by the continuous and dashed lines, respectively. (b) The stationary limit surfaces of two identical counter-rotating extreme KN black holes with $M>0$.}
\label{Stationarysurfaces}\end{figure}
\begin{figure}[ht]
\begin{minipage}{0.49\linewidth}
\centering
\includegraphics[width=4.25cm,height=5cm]{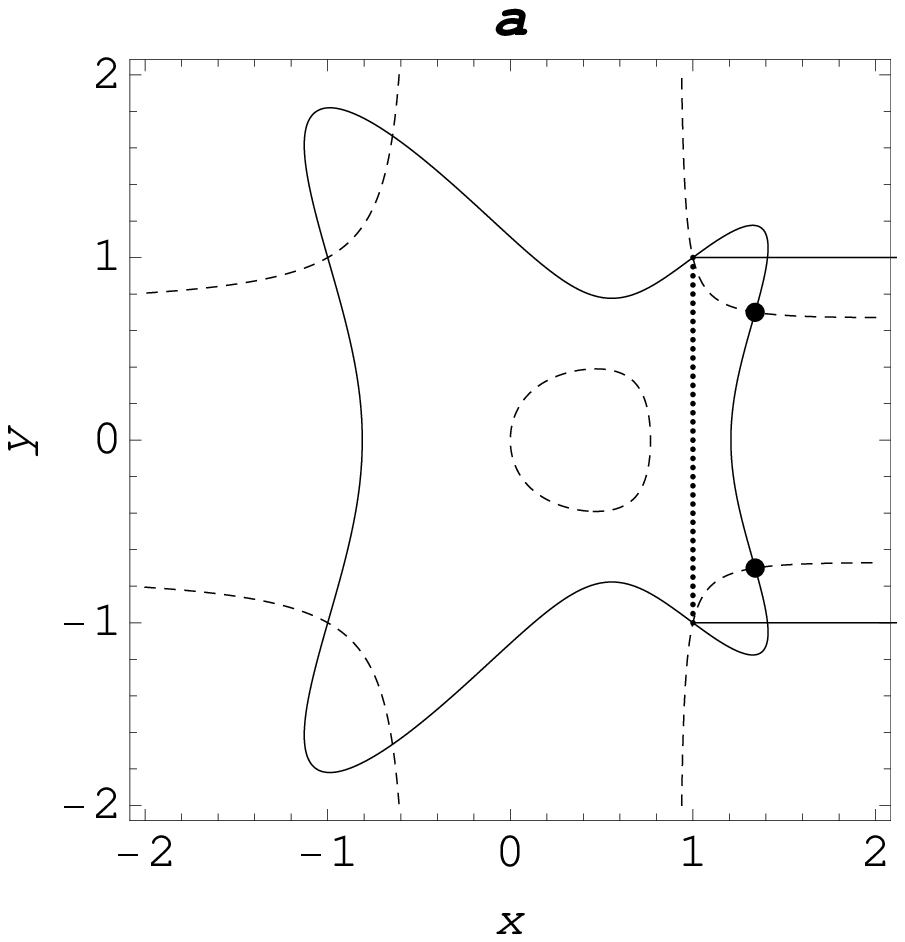}
%\center{a)} opcional
\end{minipage}
\begin{minipage}{0.49\linewidth}
\centering
\includegraphics[width=4.25cm,height=5cm]{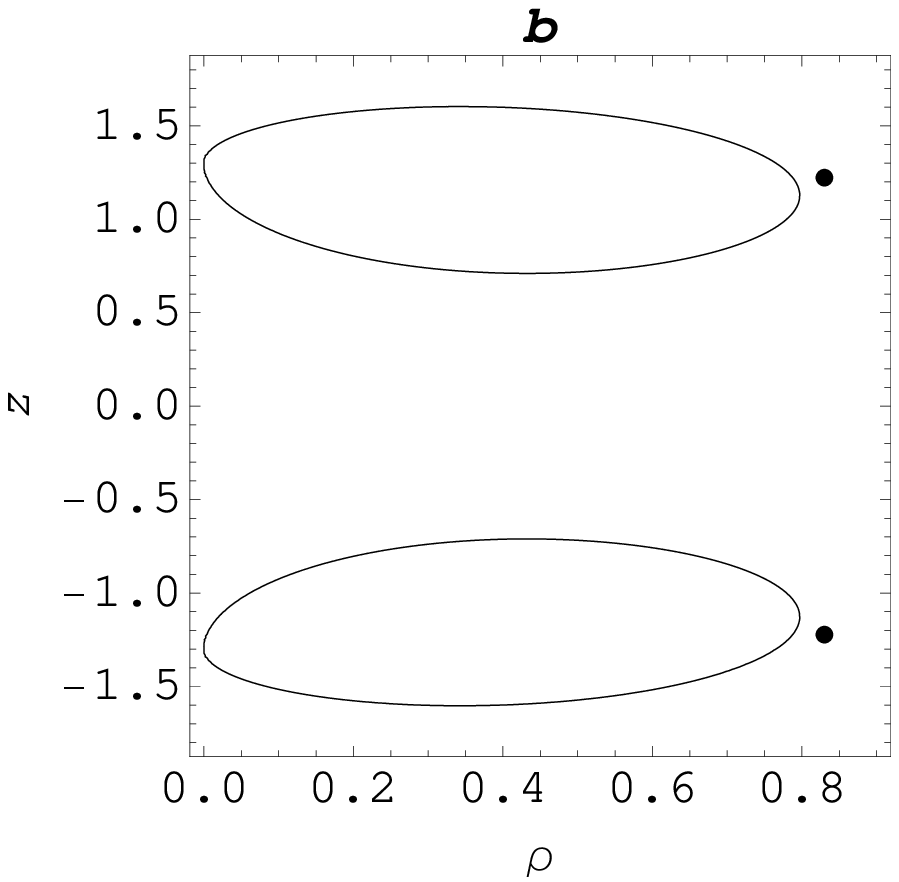}
%\center{b)} opcional
\end{minipage}
\caption{(a) Singularities located at $x=1.34$, $y=\pm 0.7$, for the values $M=-1$, $Q_{E}=-Q_{B}=-0.1$, $J=0.3$, and $\alpha \simeq 1.3$. (b) Ring singularities off the axis for $M<0$ and located at $\rho\simeq 0.83$, $z\simeq \pm 1.23$. The small displacement of the ring singularities with respect to their corresponding ergosurface is due to the presence of the electromagnetic charge.}
\label{Stationarysurfacesirregular1}\end{figure}

Additionally, the easiest analytical proof on the regularity of the solution can be made in the extreme DRN sector, since the curves defined by Eq.(\ref{curves}) are now reduced to a geometric locus of two straight lines whose intersection forms an angle of $\theta=\arctan[\alpha/\sqrt{\alpha^{2}-M^{2}}]$,
\be F_{R}\equiv F_{DRN}=\left(x+\frac{M}{\alpha}\right)^{2}-\left(1-\frac{M^{2}}{\alpha^{2}}\right)y^{2}=0,\ee

\noi where the straight lines are given by
\be y=\pm \frac{1}{\sqrt{1-M^{2}/\alpha^{2}}}\left(x + \frac{M}{\alpha}\right).\label{asymtotes}\ee

The conditions $x=1$ and $|y|<1$ are enough to prove that both straight lines cross inside the region $x>1$, $|y|<1$, forming singular surfaces off the axis (see Fig. 4),
\bea \begin{split} \frac{1}{\sqrt{1-M^{2}/\alpha^{2}}}\left(1 + \frac{M}{\alpha}\right)&<1, \quad \Rightarrow M<0, \\
-\frac{1}{\sqrt{1-M^{2}/\alpha^{2}}}\left(1 + \frac{M}{\alpha}\right)&> -1, \quad \Rightarrow M<0. \end{split}\eea

Moreover, the conditions $x=1$ and $|y|>1$ are sufficient to avoid the cross inside of such region,
and there exist no singular surfaces off the axis,
\bea \begin{split} \frac{1}{\sqrt{1-M^{2}/\alpha^{2}}}\left(1 + \frac{M}{\alpha}\right)&>1, \quad \Rightarrow M>0, \\
-\frac{1}{\sqrt{1-M^{2}/\alpha^{2}}}\left(1 + \frac{M}{\alpha}\right)&< -1, \quad \Rightarrow M>0. \end{split}\eea
\begin{figure}[ht]
\begin{minipage}{0.49\linewidth}
\centering
\includegraphics[width=4.25cm,height=5cm]{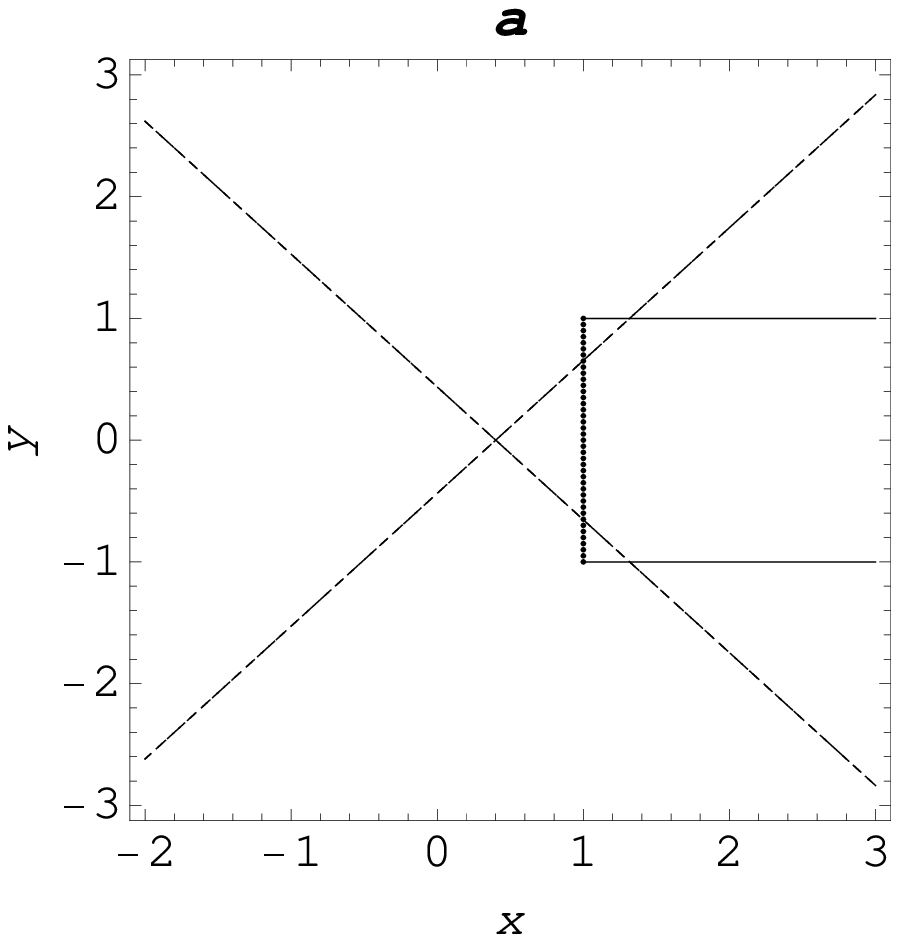}
%\center{a)} opcional
\end{minipage}
\begin{minipage}{0.49\linewidth}
\centering
\includegraphics[width=4.25cm,height=5cm]{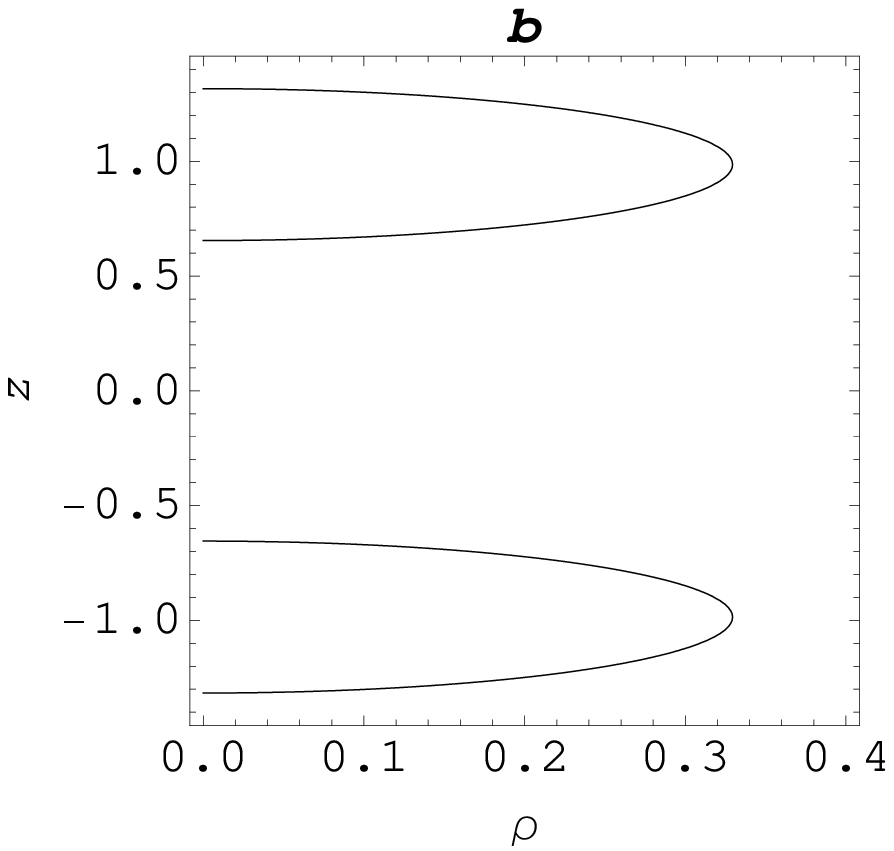}
%\center{b)} opcional
\end{minipage}
\caption{(a) Crossing inside the region $x>1$, $|y|<1$ due to $M<0$, for the values $\alpha=1$, $M=-0.4$. (b) Emergence of singular surfaces if $M<0$ in the DRN sector.}
\label{Stationarysurfacesirregular2}\end{figure}

To conclude, we should mention that our description can be reduced to the well-known limits as the vacuum and electrostatic ones \cite{Varzugin, Varzugin1}. In the extreme limit presented, it is not trivial to derive relations between the parameters and it remains as a future work to be analyzed. Particularly it would be also quite interesting to accomplish a deeper analysis of the inequalities between struts and Komar physical quantities, provided and discussed by Gabach Clement \cite{Maria}.

\section*{ACKNOWLEDGEMENTS}
This work was supported by CONACyT Grant No. 166041F3 and by CONACyT fellowship with CVU No. 173252. C. L. acknowledges support by the DFG Research Training Group 1620 ``Models of Gravity'', and by the center of excellence QUEST.


\begin{thebibliography}{99}
\bibitem{KramerNeugebauer}{D. Kramer and G. Neugebauer, Phys. Lett. A \textbf{75}, 259 (1980).}

\bibitem{Neugebauer}{G. Neugebauer and J. Hennig, Gen. Relativ. Gravit. \textbf{41}, 2113 (2009).}

\bibitem{Hoenselaers}{C. Hoenselaers, Prog. Theor. Phys. \textbf{72}, 761 (1984).}

\bibitem{MRS}{V. S. Manko, E. Ruiz, and J. D. Sanabria-G\'omez, Classical Quantum Gravity \textbf{17},
3881 (2000).}

\bibitem{Israel}{W. Israel, Phys. Rev. D \textbf{15}, 935 (1977).}

\bibitem{MRR}{V. S. Manko, R. I. Rabadan, and E. Ruiz, Classical Quantum Gravity  \textbf{30}, 145005 (2013).}

\bibitem{Parker}{L. Parker, R. Rufinni, and D. Wilkins, Phys. Rev. D \textbf{7}, 2874 (1973).}

\bibitem{BM}{N. Breton and V. S. Manko, Classical Quantum Gravity \textbf{12}, 1969 (1995).}

\bibitem{Smarr}{L. Smarr, Phys. Rev. Lett. \textbf{30}, 71 (1973).}

\bibitem{Varzugin}{G. G. Varzugin, Theor. Math. Phys. \textbf{116}, 1024 (1998).}

\bibitem{ICM}{I. Cabrera-Munguia, C. L\"{a}mmerzahl, and A. Mac\'ias, Classical Quantum Gravity \textbf{30},
175020 (2013).}

\bibitem{Komar}{A. Komar, Phys. Rev. \textbf{113}, 934 (1959).}

\bibitem{Tomimatsu1}{A. Tomimatsu, Prog. Theor. Phys. \textbf{72}, 73 (1984).}

\bibitem{EMR}{F. J. Ernst, V. S. Manko, and E. Ruiz, Classical Quantum Gravity \textbf{23}, 4945 (2006).}

\bibitem{Ernst}{F. J. Ernst, Phys. Rev. \textbf{168}, 1415 (1968).}

\bibitem{Papapetrou}{A. Papapetrou, Ann. Phys. (Leipzig) \textbf{447}, 309 (1953).}

\bibitem{Sibgatullin}{N. R. Sibgatullin, \emph{Oscillations and Waves in Strong Gravitational and
Electromagnetic Fields} (Springer-Verlag, Berlin, 1991); V. S. Manko and N. R. Sibgatullin, Classical Quantum Gravity \textbf{10}, 1383 (1993).}

\bibitem{RMJ}{E. Ruiz, V. S. Manko, and J. Mart\'in, Phys. Rev. D \textbf{51}, 4192 (1995).}

\bibitem{NUT}{E. Newman, L. Tamburino, and T. Unti,  J. Math. Phys. (N.Y.) \textbf{4}, 915 (1963).}

\bibitem{Varzugin1}{G. G. Varzugin and A. S. Chystiakov, Classical Quantum Gravity \textbf{19}, 4553 (2002).}

\bibitem{Weinstein}{G. Weinstein, Commun. Pure Appl. Math. \textbf{43}, 903 (1990).}

\bibitem{Tomimatsu0}{A. Tomimatsu, Prog. Theor. Phys. \textbf{70}, 385 (1983).}

\bibitem{Kinnersley}{W. Kinnersley, J. Math. Phys. (N.Y.) \textbf{18}, 1529 (1977).}

\bibitem{KCH}{W. Kinnersley and D. M. Chitre, J. Math. Phys. (N.Y.) \textbf{19}, 2037 (1978).}

\bibitem{Emparan}{R. Emparan, Phys. Rev. D \textbf{61}, 104009 (2000).}

\bibitem{Cabrera}{I. Cabrera-Munguia, V.S Manko, and E. Ruiz, Gen. Relativ. Gravit. \textbf{43}, 1593 (2011).}

\bibitem{SchoenYau1}{R. Schoen and S.-T. Yau, Commun. Math. Phys. \textbf{65}, 45 (1979). }

\bibitem{SchoenYau2}{R. Schoen and S.-T. Yau, Commun. Math. Phys. \textbf{79}, 231 (1981). }

\bibitem{Maria}{M. E. Gabach Clement, Classical Quantum Gravity \textbf{29}, 165008 (2012).}

\end{thebibliography}
\end{document}